%Paper: cond-mat/9410103
%From: privman@albert.phy.clarkson.edu (Vladimir Privman)
%Date: Thu, 27 Oct 1994 13:01:11 -0400

% The paper is in plain TeX  and it can be
% TeXed directly as provided. No input files,
% etc., are needed. The FIGURES ARE NOT INCLUDED.
% They can be e-mailed as PostScript files on request,
% addressed to:   privman@albert.phy.clarkson.edu

\def\NI{\noindent}

\long\def\UN#1{$\underline{
{\vphantom{\hbox{#1}}}\smash{\hbox{#1}}}$}

\magnification=\magstep 1
\overfullrule=0pt
\hfuzz=16pt
\voffset=0.0 true in
\vsize=8.8 true in
\baselineskip 16pt
\parskip 2pt
\hoffset=0.1 true in
\hsize=6.3 true in
\nopagenumbers
\pageno=1
\footline={\hfil -- {\folio} -- \hfil}

\centerline{\UN{\bf Multiparticle Reactions with
Spatial Anisotropy}}

\vskip 0.2in

\centerline{{\bf Vladimir Privman},$^a$\ \
{\bf Enrique Burgos}$^b$\ \ and\ \
{\bf Marcelo D.~Grynberg}$^b$}

\vskip 0.1in

\centerline{\sl $^a$Department of Physics,
Clarkson University, Potsdam,
New York 13699--5820, USA}

\centerline{\sl $^b$Departamento de F\'{\i}sica,
Comisi\'on Nacional de Energ\'{\i}a At\'omica,}
\centerline {\sl Avenida del Libertador
8250 (1429) Buenos Aires, Argentina}

\vskip 0.2in

\centerline{\bf ABSTRACT}

We study the effect of anisotropic
diffusion on the one-dimensional
annihilation reaction $k$A$\to$inert
with partial reaction probabilities when
hard-core particles meet in groups
of $k$ nearest neighbors.
Based on scaling arguments,
mean field approaches and
random walk considerations we argue that
the spatial
anisotropy introduces no appreciable changes as
compared to the isotropic case. Our conjectures
are supported by numerical simulations for slow reaction rates,
for $k=2$ and $4$.

\vskip 0.1in

{\bf PACS}: 68.10.Jy, 05.70.Ln, 82.20.Mj

\hrule
\vskip 0.2in

\NI {\bf 1. INTRODUCTION}

In this work we consider the effect of
anisotropy on one-dimensional annihilation reaction
processes $k$A$\to$inert, with diffusion
and hard-core particle interactions.
Recent interest in stochastic dynamics
of low-dimensional many-body systems has been
largely due to the fact that mean-field
``rate equation'' description applies in many
instances not only in high dimensions, $D$, but
also down to $D=1$ or $D=2$. Thus,
in order to observe fluctuation-dominated behavior,
one has to consider
low-dimensional systems. Specifically, in $1D$, a
host of new exact solutions,
asymptotically exact scaling
arguments, and equivalence to other more traditional
$1D$ many-body systems, for instance, Heisenberg spin
chains, have been reported [1-15].
Some of these results have experimental relevance
[16-17].

Recent work by Janowsky [18] provided an
indication that spatial anisotropy in diffusion rates
(i.e., unequal  hopping probabilities to the left and
to the right on the $1D$ line) modifies the fluctuation
behavior of the annihilation reaction
A$+$B$\to$inert. In particular, the exponent of
the large-time, $t$, power-law decay of
the particle concentration,

$$c(t) \sim t^{-\alpha}\;\; , \eqno(1.1)$$

\NI which applies provided the
initial A- and B-densities were equal, was changed from
$\alpha = 1/4$ to $1/3$. Janowsky's
observation was largely numerical, and we are not
aware of any analytical or phenomenological explanation
available in the literature.

In order to gain insight into the origins of
the asymptotic behavior for
anisotropic diffusion, we propose,
in the present work, to consider reactions
$k$A$\to$inert, i.e., $k$-particle annihilation on the line.
The reason for favoring these reactions is that
for $k>3$ the asymptotic large-time behavior
in the isotropic case is mean-field [3,19].
There are several phenomenological arguments for
this mean-field behavior,
as well as for the fluctuation-dominated behavior
for $k=2$ \ (while the case $k=3$ is marginal).
We survey these arguments in Section~2 and propose extensions
to the anisotropic case.

We then report, in Section~3, numerical results for $k=2$ and 4.
Both analytical  considerations and numerical results
suggest  that spatial anisotropy has little effect on
the overall fluctuation vs.\ mean-field regimes for these
reactions. Large-scale numerical simulations indicate
that details of the
variation of the particle density with
time are similar and
allow verification of
the exponent predictions.
A brief summary is given is Section~4.

\vskip 0.2in

\NI {\bf 2. PHENOMENOLOGICAL CONSIDERATIONS}

The early arguments [20] yielding
the fluctuation-dominated, diffusive
behavior, for instance, for the
reaction $2$A$\to$inert, were simplistic but powerful.
Thus the line of the argument went something like this:
the average particle density
$c(t)$ at time $t$ implies the average
separation of order $1/c$ (in $1D$). Given the
time $t$, the diffusion constant (of a single particle)
${\cal D}$, and this separation, the
only dimensionless combination is ${\cal D}tc^2$, which
implies $c \sim ({\cal D}t)^{-1/2}$. This
approach was further refined in [3]. Indeed, assuming
that the fluctuation behavior is universal, i.e.,
that the initial density $\rho = c(0)$ is
``forgotten'' at large times, one can
write down the large-time scaling relation (given
here for $1D$ only),

$$ c \simeq \rho F( {\cal D}t \rho^2 ) \;\; , \eqno(2.1) $$

\NI where the scaling function $F$ is 1
for small arguments but must be power-law for
large arguments such that the $\rho$-dependence
is canceled out. The resulting
prediction is

$$ c(t) \simeq {{\rm universal\ constant}
\over \sqrt{{\cal D}t} } \;\; . \eqno(2.2) $$

These arguments are valid provided
the initial configuration is random (has no
correlations, so that the only
length scale is $1/\rho$). Their advantage is that
they can be extended to $D>1$ and,
occasionally, to some more complicated reactions
[3,21]. When compared to the predictions of
the mean-field rate equations, they
yield the upper critical dimension values.
For instance for the $k=2$ reaction
considered earlier, one gets $\alpha = D/2$ and,
when compared to the mean-field value
$\alpha=1$, the upper critical dimension
is identified as $D=2$. The disadvantage of
such considerations is that they do not yield
any associated approximation scheme for
more detailed calculations.

We will term the preceding arguments
``the scaling approach.'' Making diffusion
anisotropic modifies the diffusion constant
and also introduces a new dimensional
quantity, the drift velocity. Its effect on
the scaling-approach predictions is not
clear in general. We note, however, that the
``scaling approach'' was mainly used for
the simplest two-body reactions for which the
drift should not be important. Indeed,
in the fluctuation-dominated regime
the reaction rate eventually renormalizes to the
``fast-reaction'' (so-called ``diffusion-limited'')
limit. The interparticle
distributions are believed to approach
diffusion-dominated forms [13,22-24], with
effects of other particle interactions
(such as hard-core, etc.) being ``renormalized
away.'' In the drifting reference frame, therefore,
the reaction will be the same with
slowed-down diffusion. Thus the results of
[18] for the reaction A$+$B$\to$inert,
which indicate that anisotropic diffusion
can modify the fluctuation-dominated behavior, were surprising and
they are still largely open to
interpretation.

Another approach was developed in [19].
Assuming that mean-field applies, one uses the
uncorrelated form of the interparticle
distance distribution to make specific
predictions of the reaction event rates.
The result is a mean-field rate equation.
When combined with phenomenological
arguments similar to those described earlier, i.e.,
diffusive time-scale estimates, this approach
can predict, for instance, that the
$k>3$ annihilation reactions are
mean-field in $1D$ (for isotropic diffusion), while
the $k=3$ reactions are marginal, which was
recently confirmed numerically and
analytically [25], and the $k=2$ case
is fluctuation-dominated. Similar mean-field
approximations for other $1D$ correlation
quantities were used to study more
complicated mostly two-body reactions [23-24]
in $1D$. The advantage of these
approaches is that they are associated
with specific calculational schemes which yield
results in the mean-field approximation
by typically providing a rate equation for the
density and other quantities.

The disadvantage of these mean-field
based approximations is that they cannot be easily
extended to $D>1$ because explicit results
for uncorrelated particle systems are used
[19], or objects which are only appropriate in
$1D$ are considered [23-24]. (There are, however, numerous
other mean-field formulations not limited to $1D$.)
We note that the approach of [19] for instance,
only requires that for fast diffusion
(i.e., assuming negligibly slow reaction rate),
the particles become uncorrelated. As in [19], we
consider here particles diffusing on the line and
reacting in groups of $k$ on encounters.
However, if less than $k$ particles meet,
they interact as hard-core objects.
In Figure~1 we show a snapshot of the
stochastic evolution
of such an anisotropic process for $k=4$. Further details
will be given in Section~3.

An important observation is that
making the diffusion anisotropic does not change the
property of the hard-core particle system
to loose correlation at large times [4],
due to diffusion only (disregarding the reaction).
This would suggest that the
approximations of [19] should remain valid.
The behavior for $k>3$ should be
mean-field for anisotropic hopping,
with $k=3$ marginal,
while $k=2$ is non-mean-field
(although the arguments of [19] cannot
predict the exponent $\alpha$).

Finally, we consider yet another line of
argument based on more local, ``relative
coordinate'' considerations.
These ideas are similar to those advanced in
the exact asymptotic-limit studies of two-particle
reactions [11,13,22]. For illustration, let us
consider the isotropic-case $k$-particle
annihilation in $1D$. We note that when $k$
particles needed for reaction end up as
a close group due to diffusion, the ``memory''
of this event will be ``washed away''
by diffusion provided the random walk in the
$k-1$ relative distances is non-recurrent.
Only for the 1- or 2-dimensional
relative-coordinate walk, will the encounter
be likely repeated with the same particles
(assuming that the system is already quite dilute).
Thus, local fluctuations are
likely to be less important for $k-1>2$,
which is consistent with the earlier
identification of the borderline value $k=3$
which separates the mean-field and
fluctuation-dominated regimes.

The advantage of the ``relative coordinate''
argument is that it is local and
therefore it can be easily applied in $D>1$, etc.
However, the ``locality'' is also
the disadvantage. Indeed, the fluctuation-dominated
behavior is generally
considered to be a genuinely many-body effect.
Few-particle arguments can at
best provide hints on the various
regimes of behavior.

Since the hard-core constraint
is short-range, it should have no effect when
the particles are well separated, in the dilute limit.
Therefore, the recurrence or nonrecurrence property
in the relative coordinates will be the
same for isotropic or anisotropic diffusion
(as long as no finite-size effects are
present, i.e., as long as the particles cannot
interact ``around'' the system for
periodic boundary conditions, for instance).
In fact, for $k=2$ we anticipate small or
no changes as compared to the isotropic case.
Our numerical results (next section) have verified this
expectation and also indicated a similar property for $k=4$.

\vskip 0.2in

\NI {\bf 3. NUMERICAL RESULTS}

The specific details of the hard-core interaction
and annihilation reactions, used in our
numerical simulations, are given below. We note, however,
that the conclusions regarding the asymptotic large-time
behavior are generally expected to be universal. The numerical results
reported were ``large-scale,'' computer-resource demanding, which was the
main reason for considering only the cases $k=2$ (non-mean-field)
and $k=4$. The latter provides the
simplest example of the ``clean'' mean-field
behavior already well-studied for the isotropic diffusion [19].

At time $t=0$, each site of the one-dimensional
lattice with periodic boundary conditions
is occupied with probability $c (0)=\rho$
by identical hard-core particles. The particles perform
a biased random walk between nearest-neighbor
lattice sites.
At a given time $t$ one of the $N(t)$ particles
present in the system is picked at random.
Let $j$ denote its lattice site location.
This particle hops to the site $j+1 \;$
(or $j-1 $) with probability $h \;$ (or $1-h $) provided
the target site is vacant, otherwise it remains in place.
After each successful hopping attempt, the active
particle
can annihilate with probability $q \leq 1$ with
$k-1 $ consecutive particles located {\it in the direction of the
hopping}, i.e., on sites
$j+2,  \ldots,   j+k \; $ (or $j-2, \ldots, j-k $),
provided of course that all the
$k-1 $ corresponding neighboring sites were
already occupied.
If the target site $j+1 \;$ (or $j-1 $)
was already occupied, so that the hopping event did
not take place, the active particle at $j$
may still annihilate (with probability $q $)
with $k-1 $
particles at sites
$j+1,  \ldots, j+k-1 \;$ (or $j-1, \ldots,  j-k+1 $),
provided they were all occupied. This rule, involving
annihilation with particles
in the direction of the hopping attempt,
successful or unsuccessful, introduces
correlation between hopping and reaction
which has some similarity with the actual
chemical reactions in $D>1$; see [19].

Each successful annihilation reduces the
total number of particles present in the
system by $k $. The numerical procedure allows for
$N$ such hopping with reaction attempts per each time unit.
Thus, only after $N$ attempts the time is increased by 1, and the
particle number $N(t+1)$ is recalculated. This methodology
is particularly efficient for dilute regimes and
sets up a well-defined time scale
since on average each particle is selected once
per unit time. Of course, the results are only meaningful for
sufficiently large system sizes so that $N(t) \gg k$ holds
for all times $t$ studied. (Strictly speaking, low reaction probability
is also required when particle density is of order 1.)

Our Monte Carlo simulations confirm
the general theoretical expectations for $k=2 $ and $4 $.
Results for the density, $c(t)$, are illustrated in Figure~2.
In order to investigate any changes due to anisotropy of hopping
in the large-time asymptotic regime, we studied the case $h=0.5 $,
compared to the maximal anisotropy, $h=1$. In fact,
large-scale simulations up to
$10^8 $ time steps, starting with random, homogeneous initial particle
distribution with densities $\rho = 0.25,  0.5 $ and $0.8 $,
indicate that no such differences
are present, even for rather slow reactions,
$q \ll 1 $ (our numerical values were as low as 0.001).
Moreover, the evolution
of the particle concentration
for $h=1 $ closely follows the density
of the isotropic case for all times. For
{\it instantaneous\/} reactions ($q=1 $) of
{\it two\/} particles this result was shown to be
rigorous [15,26].

For slow annihilation rates the
case $k=2 $ exhibits a regime of the mean-field
like behavior ($ c \sim t^{-1} $), followed by a crossover to the
fluctuating-dominated asymptotic power law ($c \sim t^{-1/2} $).
This is illustrated by the $q=0.01$ data in Figure~2(a).
These data were averaged
over 200 independent Monte Carlo runs for periodic lattices
of $6 \times 10^4$  sites. Although finite size effects were
not severe,
rather large lattice sizes were necessary
to explore large-time behavior while keeping the particle
number $N(t)$ large, as discussed earlier. The mean-field
regime became more pronounced as $q$ decreased. For instance,
for $q=0.001$ (not shown here)
it extended over more than three decades
in the $t$ variable.

However, the asymptotic behavior of the $k=2$
reaction-diffusion system is always
fluctuation-dominated, $\alpha = 1/2$, and the anisotropy
introduces no detectable changes in the density. Interestingly,
the initial density is ``forgotten'' before the onset of the
mean-field-regime behavior
(and of course the initial short-range correlations
are completely ``washed away'' in
the fluctuation-dominated asymptotic limit).
This property is shared by the $k=4$ results; see Figure~2(b).

For $k=4 $ the asymptotic behavior was found to be
mean-field for both
isotropic and anisotropic hopping. The density variation
of the fully anisotropic case ($h=1$) was in close numerical
agreement with the results obtained for isotropic hopping ($h=0.5 $)
and therefore it can be described accurately by the
mean-field rate-equation calculation scheme
given in [19].
Since the asymptotic particle concentration decay
is now slower ($ c \sim t^{-1/3} $), smaller lattice
sizes can be used to
investigate the behavior for large times.
For instance, the $q=0.001$ data shown in Figure~2(b) represents
average over 200
runs, for a periodic lattice of
$2 \times 10^3 $ sites.

\vskip 0.2in

\NI {\bf 4. DISCUSSION}

Generally, our simulations have confirmed the
phenomenological considerations regarding the
asymptotic particle density, presented in Section~2.
However, the observation that the data
for the isotropic and anisotropic cases are so close
numerically for all times and $q$-values has not been
explained adequately.

Our dynamical rules introduce correlations
between hopping and reaction, however
they are more appropriate to describe
actual chemical systems. Specifically,
the partial reaction probability rates
$q <  1\,$  might be interpreted as the result
of an effective potential that particles must
overcome in order to annihilate.
Thus, collisions between particles (due to diffusion)
should promote the particle
cluster to go over the reaction energy barrier.
Although these correlated processes are
less well described by mean-field
calculation schemes, for $k > 3\,$
the use of this methodology is still justified
within the fast-diffusion regime ($q \ll 1\,$)
where the role of such correlations becomes
irrelevant both in the isotropic and anisotropic case.
It is worth pointing out however, that
for $k=2\,$ and $q=1\,$, the decoupling between
annihilation and diffusion allows for an exact solution
of the macroscopic particle concentration
which is independent of the hopping anisotropy. To
the best of our knowledge, no way of solving the case
$k=2\,$ is presently known either with
partial annihilation rates or hopping-reaction
correlations.

Finally, our numerical results were only for
systems with periodic boundary conditions.
It is well established that
{\it pair correlations\/} in hard-core
(nonreacting) particle systems with {\it
anisotropic diffusion\/} are extremely sensitive
to boundary conditions. It is also expected that
anisotropic diffusion might introduce significant
changes in the form of {\it unequal-time} correlation
functions (not studied in our present simulations).
Specifically, for $k=2 $  and $q=1 $
exact analyses are feasible using
fermionic techniques. Thus, we hope that the
present work will set the stage
for further studies of the anisotropic
multiparticle reactions.

\vskip 0.2in

{\bf ACKNOWLEDGMENTS}

V.P.~wishes to thank Professor J.L.~Lebowitz
for helpful discussions. E.B.~and
M.D.G.~gratefully acknowledge the financial
support of the Consejo Nacional de
Investigaciones Cient\'{\i}ficas y
T\'ecnicas of Argentina (CONICET), Fundaci\'on
Sauberan and Fundaci\'on Antorchas. M.D.G.~is grateful
to the Department of Theoretical Physics,
University of Oxford, UK, for the allocation of
computing facilities.

\vskip 0.2in

\centerline{\bf REFERENCES}{\frenchspacing \baselineskip 15pt

\item{[1]} M. Bramson and D. Griffeath,
Ann. Prob. {\bf 8}, 183 (1980).

\item{[2]} D.C. Torney and H.M. McConnell,
J. Phys. Chem. {\bf 87}, 1941 (1983).

\item{[3]} K. Kang, P. Meakin, J.H. Oh and S. Redner,
J. Phys. A {\bf 17}, L665 (1984).

\item{[4]} T. Liggett,
{\sl Interacting Particle Systems\/}
(Springer-Verlag, New York, 1985).

\item{[5]} A.A. Lushnikov, Phys. Lett. A {\bf 120},
135 (1987).

\item{[6]} V. Kuzovkov and E. Kotomin, Rep. Prog. Phys.
{\bf 51}, 1479 (1988).

\item{[7]} J.L. Spouge, Phys. Rev. Lett. {\bf 60},
871 (1988).

\item{[8]} D.J. Balding and N.J.B.
Green, Phys. Rev. A {\bf 40}, 4585
(1989).

\item{[9]} J.G. Amar and F. Family, Phys. Rev. A {\bf 41},
3258 (1990).

\item{[10]} Z. Racz, Phys. Rev. Lett. {\bf 55},
1707 (1985).

\item{[11]} M. Bramson and
J.L. Lebowitz, Phys. Rev. Lett. {\bf 61},
2397 (1988).

\item{[13]} D. ben--Avraham, M.A.
Burschka and C.R. Doering, J. Stat. Phys. {\bf 60},
695 (1990).

\item{[14]} V. Privman, J. Stat. Phys.
{\bf 69}, 629 (1992).

\item{[15]} The equivalence of
adsorption-desorption processes, reaction-diffusion systems
and spin-chain dynamics has been explored by M.D.
Grynberg, T.J. Newman  and R.B. Stinchcombe,
Phys. Rev. E {\bf 50}, 957 (1994);
M.D. Grynberg and R.B. Stinchcombe, Phys. Rev. E {\bf 49},
R23 (1994); M. Barma, M.D. Grynberg and R.B. Stinchcombe,
Phys. Rev. Lett. {\bf 70}, 1033 (1993);
R.B. Stinchcombe, M.D. Grynberg and
M. Barma, Phys. Rev. E {\bf 47}, 4018 (1993);
G.M. Sch\"utz,  J. Stat. Phys., in print (1994).

\item{[16]} R. Kopelman, C.S. Li and Z.--Y. Shi,
J. Luminescence {\bf 45}, 40 (1990).

\item{[17]} R. Kroon, H. Fleurent and R. Sprik, Phys. Rev.
 E {\bf 47}, 2462 (1993).

\item{[18]} S.A. Janowsky, preprint.

\item{[19]} V. Privman and M.D. Grynberg, J. Phys.
A {\bf 25}, 6575 (1992).

\item{[20]} K. Kang and S. Redner,
Phys. Rev. Lett. {\bf 52}, 955 (1984); Phys. Rev. A {\bf 30},
2833 (1984).

\item{[21]} S. Cornell, M. Droz and
B. Chopard, Phys. Rev.
A {\bf 44}, 4826 (1991).

\item{[22]} C.R. Doering and D. ben--Avraham, Phys. Rev.
A {\bf 38}, 3035 (1988).

\item{[23]} J.C. Lin, C.R.
Doering and D. ben--Avraham, Chem. Phys.
{\bf 146}, 355 (1990).

\item{[24]} V. Privman, C.R. Doering
and H.L. Frisch, Phys. Rev. E {\bf 48}, 846 (1993).

\item{[25]} D. ben--Avraham, Phys. Rev. Lett.
{\bf 71}, 3733 (1993); P.L. Krapivsky, Phys. Rev. E {\bf 49},
3223 (1994); B.P. Lee, J. Phys.
A {\bf 27}, 2533 (1994).

\item{[26]} V. Privman,
J. Stat. Phys. {\bf 72}, 845  (1993).

}

\vskip 0.2in

\centerline{\bf FIGURE CAPTIONS}

\noindent\hang{\bf Figure~1.}\
Snapshot of (a part of) the stochastic evolution for the fully
anisotropic reaction-diffusion
system 4A$\to$inert. Details of the numerical procedure are
described in Section~3; specifically, the case illustrated was for
the reaction probability $q=0.1$.

\noindent\hang{\bf Figure~2.}\ Macroscopic density
$c(t)$ obtained numerically starting from the random particle
distribution with the initial
concentrations $\rho = 0.25,  0.5 $ and $0.8 $.
The open symbols correspond to the fully anisotropic
hopping ($h=1 $). The small dots joined by solid
lines denote the isotropic-hopping results ($h=0.5 $).
(a) Data for $k=2 $ and reaction probability
$q = 0.01 $, averaged over $200$ Monte Carlo runs, for a
periodic lattice of $6 \times 10^4 $ sites.
(b) Results for $k=4 $, $q = 0.001 $, for
$2 \times 10^3 $ lattice sites, averaged over
$200$ runs. The dashed lines
indicate the asymptotic slopes corresponding
to the predicted power-law behavior described in the text.

\bye